\documentclass[aps,pra,reprint,floatfix,showpacs,nofootinbib]{revtex4-1}
\usepackage{amssymb,amsmath,graphicx,color}

\begin{document}

\title{Observation and Analysis of Resonant Coupling Between Nearly Degenerate Levels of the $2 \, ^1 \Sigma _g^+$ and $1 \, ^1 \Pi _g$ States of Ultracold $^{85}\text{Rb}_2$}

\author{R. Carollo}
\author{M. A. Bellos}
\author{D. Rahmlow}
\author{J. Banerjee}
\author{E.E. Eyler}
\author{P.L. Gould}
\author{W.C. Stwalley}
\affiliation{Department of Physics, University of Connecticut, Storrs, CT 06269}

\date{\today}


\begin{abstract}
We report on the anomalously high line strength of a single rotational level in the ultracold photoassociation of two $^{85}$Rb atoms to form $^{85}$Rb$_2$. The $v' = 111$, $J' = 5$ level belongs to the Hund's case~(c) $ 2 \, (0_g^+)$ state, which correlates to the Hund's case~(a) $2 \, ^1 \Sigma _g^+$ state. Its strength is caused by coupling with a very near-resonant long-range state. The long-range component is the energetically degenerate $v' = 155$, $J' = 5$ level of the case~(c) $2 \, (1_g)$ state, correlating to the case (a) $1 \, ^1 \Pi _g$ state. The line strength is enhanced by an order of magnitude through this coupling, relative to nearby vibrational levels and even to nearby rotational levels of the same vibrational level. This enhancement is in addition to the enhancement seen in all $J' = 3$ and 5 levels of the $ 2 \, (0_g^+)$ state due to an $l = 4$ shape resonance in the $a \,^3 \Sigma_u^+$ state continuum, which alters the distribution of levels formed by photoassociation.
\end{abstract}

\maketitle

\section{Introduction}
Ultracold molecules are currently a topic of much interest in the atomic and molecular physics community. When created through photoassociation (PA) of ultracold atoms, which is one of the simplest experimental techniques, these molecules can be used for high-resolution spectroscopy \cite{stwalley99}. For many other applications, there is strong interest in controlling the final state of the molecule. In the case of alkali-metal dimers, levels of both the $X \, ^1 \Sigma _g^+$ and $a \, ^3 \Sigma _u^+$ states are stable enough for study.  In either state, the vibrational and rotational levels populated are the primary variable that experimenters want to control. Several groups have had success in using PA to produce molecules in $v'' = 0$ of the $X \, ^1 \Sigma _{(g)}^+$ state in K$_2$ \cite{gould00}, Cs$_2$ \cite{lignier11}, LiCs \cite{weidemuller08}, NaCs \cite{wakim12}, KRb \cite{banerjee12}, and RbCs \cite{gabbanini12, bruzewicz12}. Our own group has previously formed molecules in the lowest triplet $a \, ^3 \Sigma _u^+$ state of $^{85}$Rb$_2$ in vibrational levels $v'' = 32-35$ \cite{huang06-2} and $v'' = 0$ \cite{bellos11}, and formation of $v'' = 0$ molecules was also previously reported in \cite{grimm08}.

There is a significant body of literature regarding enhancement of PA via resonant coupling, and its use is becoming important in many experiments. Resonant coupling is experimentally useful because the long-range state enhances the PA rate, while the short-range state enhances decay to desirable (typically deeply-bound) levels. Our group has previously studied several such resonant couplings in $^{85}$Rb$_2$ \cite{pechkis07} and $^{39}$K$^{85}$Rb \cite{banerjee12}. We have also discussed potential applications of these couplings and other predicted couplings in the creation of ground-state molecules \cite{stwalley10}. Other work using resonant coupling includes NaCs \cite{bigelow11} and Cs$_2$ \cite{pillet01}. Here, we demonstrate a pathway to form molecules in $v'' = 18-24$ of the $a \,^3 \Sigma_u^+$ state via a near-degenerate resonant coupling between levels of the  Hund's case (c) 2 $(0_g^+)$ state at short range and the $2 \, (1_g)$ state at long range.  The $2 \, (0_g^+)$ state correlates with the Hund's case (a) $2 \, ^1 \Sigma _g^+$ state, while the $2 \, (1_g)$ correlates to the Hund's case (a) $1 \, ^1 \Pi_g$ state.

\section{Experiment}\label{experiment}

\begin{figure}[tb]
\begin{center}
\includegraphics[width=\columnwidth]{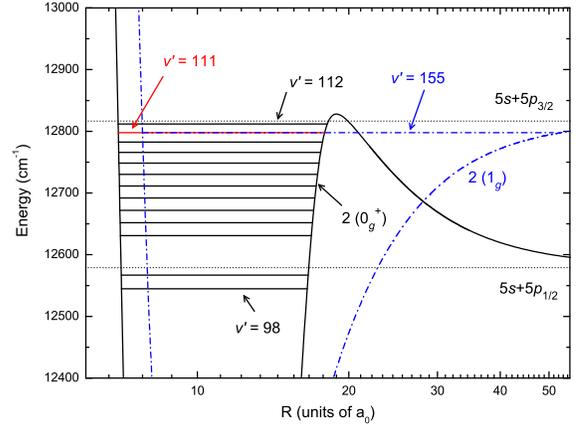}
\caption{(Color online) Potential energy curves for the $2 \, (0_g^+)$ and $2 \, (1_g)$ states as calculated by \cite{dulieu11}. The levels of the $2 \, (0_g^+)$ state observed in \cite{bellos12} are shown, as is the $v' = 155$ level of the $2 \, (1_g)$ state that is energetically near-degenerate with $v' = 111$.}
\label{levels}
\end{center}
\end{figure}

Our experimental apparatus has been described in detail in Ref. \cite{bellos12}. We begin by loading a magneto-optical trap (MOT) with $8 \times 10^7$ $^{85}$Rb atoms at a peak density of $10^{11}$ cm$^{-3}$ and a temperature of $120 \, \mu$K. We then excite free-to-bound transitions to the $2 \, (0_g^+)$ state converging to the $5s_{1/2} + 5p_{1/2}$ asymptote. It rapidly decays radiatively to form metastable molecules in the $a ^3 \Sigma_u^+$ state, which are then ionized via $1 + 1$ resonance-enhanced multiphoton ionization (REMPI) through the $2 \, ^3 \Sigma_g^+$ or $3 \, ^1 \Sigma_g^+$ states and detected by a discrete dynode multiplier. Spectroscopy of these states was previously described in Refs.~\cite{huang06, huang06-2}. Molecular ions are discriminated from atomic ions and scattered light by time-of-flight mass spectrometry.

As we showed in recent work \cite{bellos12}, the $2 \, (0_g^+)$ state supports quasibound vibrational levels behind a barrier above the $5s+5p_{1/2}$ limit. As also described in that work, the rotational distribution of these levels is affected by a ground-state $l = 4$ shape resonance that enhances the photoassociation rate to $J' = 3$ and 5 beyond the strength of lower rotational levels, and well above the expected strength for a thermal distribution. All of the levels that we have observed are below the $5s + 5p_{3/2}$ asymptote, where there are also many levels belonging to other electronic states that correlate to that asymptote. One such level, $v' = 155 \, (\pm 1)$\footnote{The vibrational numbering is somewhat uncertain, as it is based on \emph{ab initio} potentials and no complete experimental assignment is known. We believe, however, that the assignment is good to within $\pm 1$ vibrational quantum number.} of the $2 \, (1_g)$ state (shown in Fig. \ref{levels}), is nearly energetically degenerate with $v' = 111$ of the $2 \, (0_g^+)$ state.

\begin{figure}[tb]
\begin{center}
\includegraphics[width=\columnwidth]{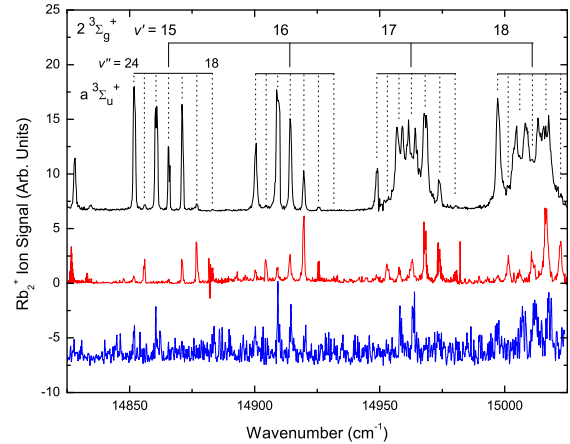}
\caption{(Color online) Three REMPI spectra of the same region, using different PA levels to form the molecules. At top (black): a spectrum taken using PA to the resonantly coupled $v' = 111, J' = 5$ level of the $2 \, (0_g^+)$ state. At center (red): a spectrum using a typical vibrational level, $v' = 107$, $J' = 3$, which is not resonantly coupled. At bottom (blue): the spectrum taken using a small satellite feature of the $v' = 111, \, J' = 5$ line, which is marked in Fig. \ref{dollloss}(b). This spectrum is the average of three scans and is enlarged by a factor of 10 for visibility. We believe that this satellite feature represents a level of coupled $2 \, (0_g^+)$ and $2 \, (1_g)$ character. Vibrational quantum numbers of the initial $a \, ^3 \Sigma_u^+$ and intermediate $2 \, ^3 \Sigma_g^+$ states are assigned as indicated.}
\label{rempi}
\end{center}
\end{figure}

When molecules in the $2 \, (0_g^+)$ state decay, they form metastable $a \, ^3 \Sigma _u^+$ molecules in levels $v'' = 18-24 $. A section of a REMPI spectrum produced from the decay of $v' = 111$ and exhibiting the $a \,^3 \Sigma_u^+$ vibrational level spacing is shown in Fig. \ref{rempi}. A similar spectrum showing molecules produced by the decay of $v' = 107$ is shown for comparison. While also strong, it has a noticeably reduced signal-to-noise ratio, indicating the $v' = 111$ level's usefulness in spectroscopic applications. Also shown is a REMPI spectrum obtained by photoassociating to a small satellite peak of the $v' = 111, J' = 5$ level. This is discussed in more detail in Section \ref{structure}, where the satellite peak is marked in the PA spectrum in Fig. \ref{dollloss}(b). For a pure $\Omega = 0$ state such as the $2 \, (0_g^+)$ state, there should be no significant hyperfine structure. Nonetheless, decay products of the satellite peak show nearly the same $a \, ^3 \Sigma _u^+$ state vibrational level distribution as the $v' = 111$, $J' = 5$ level main peak, showing that they are closely related. This unexpected substructure around $v' = 111$, $J' = 5$ is one of several indications that the long-range $2 \, (1_g)$ state is coupled with the $2 \, (0_g^+)$ state.

\section{Molecule Production}

\begin{figure}[tb]
\begin{center}
\includegraphics[width= \columnwidth]{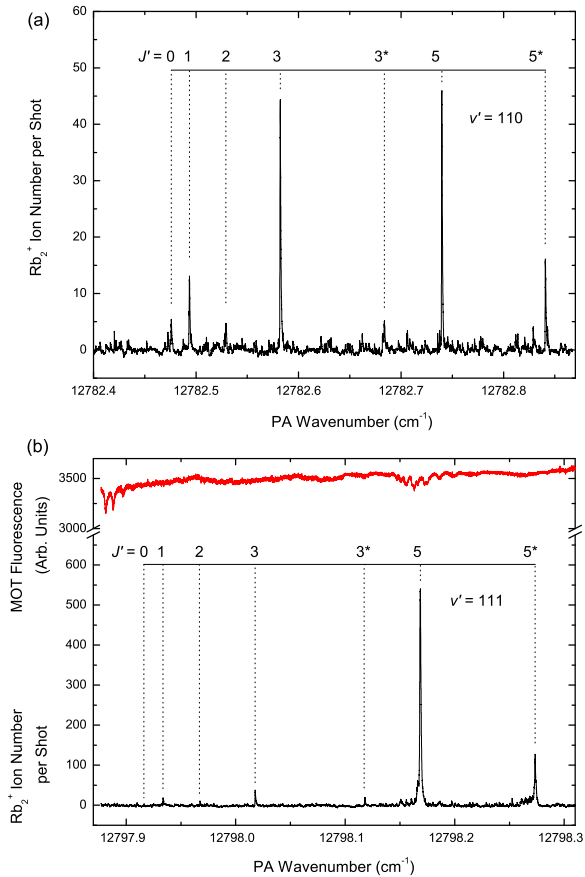}
\caption{(Color online) \textbf{(a)} The rotational spectrum of $v' = 110$ of the $2 \, (0_g^+)$ state as seen in ion detection from the $a \, ^3 \Sigma_u^+$, $v' = 20$ level, a fairly typical example of a strong PA transition to the $2 \, (0_g^+)$ state. \textbf{(b)} The spectrum of $v' = 111$ detected from the $a \, ^3 \Sigma_u^+$, $v' = 22$ level, showing the strongly-enhanced $J' = 5$ line. This spectrum has been scaled based on Lorentzian fitting of other spectra, which exhibited clipping. The resulting $J' = 3$ line strength is comparable to that of $v' = 110$ in panel (a). At the top of panel (b) is the MOT fluorescence trap loss spectrum of the long-range $2 \, (1_g)$ state. High resolution REMPI and trap loss scans of these coupled states are shown in Figs. \ref{dollloss}(b) and (a), respectively. In both panels, the lines marked with an asterisk (*) are hyperfine ``ghosts'' from the lower $F = 2$ ground-state hyperfine level of $^{85}$Rb$_2$.}
\label{pa}
\end{center}
\end{figure}

A typical PA transition to a molecular level red-detuned from the $5s + 5p_{1/2}$ or $5s + 5p_{3/2}$ asymptote can create trap loss in the MOT of several percent, and extremely strong PA lines can exceed 50\% loss. By comparison, the blue-detuned PA reported in our previous work \cite{bellos11} and in the current experiment \cite{bellos12} has never produced observable trap loss signals. ``Blue-detuned'' PA is used to denote PA to levels that are energetically above the atomic asymptote to which they correlate, and thus are quasi-bound. Other investigations of this spectral region by trap loss have also seen no evidence of blue-detuned PA \cite{heinzen94p3/2}. Nevertheless, the strongest line in this work, $v' = 111, J' = 5$, yields REMPI signals of 275 ions per REMPI shot or more under favorable conditions. The peak of this transition is clipped and therefore this value is actually a lower bound on the actual rate. Modeling the clipped line as a pure Lorentzian gives an estimated peak ion production rate of $\sim 540$ ions per shot. An unclipped spectrum has been scaled to match this peak value in Fig. \ref{pa}(b), where the line strength of the non-coupled lines ($J' = 0$--3) is seen to be comparable to the $v' = 110$ spectrum in Fig. \ref{pa}(a). We believe this similarity should exist because the ratios of the lines that are not resonantly coupled are similar, indicating that PA is not strongly affected by other factors such as the $l = 4$ ground-state shape resonance, and that the resonant coupling affects only $J' = 5$.

Using the simple method of Ref.~\cite{bellos11}, we can estimate the PA rate leading to this REMPI signal. The number of ions measured per REMPI pulse is:
\begin{equation}
N_{Rb_2^+} = N_a P_{\text{ionization}} e_{\text{d}} \text{ ,}
\end{equation}
where $e_{\text{d}}$ is the detector efficiency, $P_{\text{ionization}}$ is the probability of ionization by the REMPI pulse, and $N_a$ is the number of molecules in the relevant vibrational level of the $a \, ^3 \Sigma_u^+$ state. For a conservative estimate of the PA rate, we will assume a detector efficiency of $e_{\text{d}} = 1$, although it may be somewhat less.

The ionization probability is given by:
\begin{align}
P_{\text{ionization}} = 1-e^{-Wt} &= 1-e^{-\frac{\sigma F}{t} t} \nonumber \\
&= 1-e^{-\sigma E \lambda/ (h c \pi r^2)} \text{ .}
\end{align}
Here $W = \frac{\sigma F}{t}$ is the transition rate determined by the photoionization cross section $\sigma$ and the photon flux $F$ per unit time. In turn, the flux is $F = E \lambda/ (h c \pi r^2)$, where $E$ is the pulse energy, $\lambda$ is the pulsed laser wavelength, and $r$ is the radius of the pulsed beam in the interaction region. The pulse is assumed to have constant intensity, as the laser beam profile is highly non-Gaussian. There is little published data on ionization cross sections in Rb$_2$, especially for two-photon processes. However, the cross section of the upper REMPI photoionization step is rate limiting, as the bound-to-bound initial step is likely saturated, so we will assume that ionization acts as a one-photon process. Using the data from Ref.~\cite{samson83} and allowing for significant deviations due to the different ionization conditions, we will take $\sigma = 1^{+5}_{-0.5} \times 10^{-18} \, \mathrm{cm}^{-2}$. This results in $P_{\text{ionization}} \cong 0.235$, with lower and upper bounds of $0.125$ and $0.799$. The remaining terms are experimental parameters, with $\lambda \approx 655$ nm, $E = 5$ mJ, and $r = 1.4$ mm.

With these inputs, and using the modeled peak of $N_{Rb_2^+} = 540$, the population is $N_a \cong 2300$ molecules in the detected vibrational level, $v'' = 22$ of the $a \, ^3 \Sigma_u^+$ state, with a range of $\sim$ 675--4300 molecules. The PA rate can then be determined using
\begin{equation}
N_a (t) =  \frac{R_{\text{PA}} P_{\text{FCF}} t}{(1+t/\tau)} \text{ .}
\end{equation}
Here $R_{\text{PA}}$ is the PA rate per atom, $\tau$ is the time molecules spend in the REMPI interaction region after formation, and $P_{\text{FCF}}$ is the Franck-Condon Factor (FCF) that approximates the fraction decaying to an individual triplet level. In our system, cold molecules spend $\approx 5$ ms in the REMPI region before their velocity and the acceleration of gravity carry them out. The FCF for decay to the $v'' = 22$ level is $4.36 \times 10^{-2}$, as calculated by LEVEL 8.0 \cite{leroylevel} using the $a \, ^3 \Sigma_u^+$ potential from Ref.~\cite{tiemann10}. The estimated PA rate is thus $1.1 \times 10^{7}$ molecules per second, with a range between $3.1 \times 10^{6}$ and $2.0 \times 10^{7}$ s$^{-1}$.

An interesting comparison can be made between the PA rate calculated above from ion signals and the rate of PA at the same laser frequency implied by the observed trap loss signal. This trap loss, as seen in Figs. \ref{pa}(b) and \ref{dollloss}(a), is $\approx 2\%$ at the $2 \, (0_g^+)$, $v' = 111$ position. The trap loss is $\approx 4\%$ at the largest peaks of the $2 \, (1_g)$ state. For a MOT loaded in the presence of an extra loss mechanism, the relevant rate equation is
\begin{equation}
\frac{dN}{dt} = r_{\text{load}} - \left(\gamma + r_{\text{PA}} \right) N \text{,}
\end{equation}
where $\gamma = 1/\tau$, with $\tau$ specifying the MOT loading time without the PA beam, $r_{\text{load}}$ is the loading rate of the MOT without PA, and $r_{\text{PA}}$ is the loss rate due to PA. In a steady state, the atom number $N_0 = r_{\text{load}}/ \left(\gamma + r_{\text{PA}} \right)$. The measured values of $\tau$ and $N_0$ are $\tau = 2 \pm 1$~s and $N_0 = 8 \times 10^7$ atoms. Assuming that the PA laser is scanned slowly enough that the steady state is always maintained, the PA rate per atom is
\begin{equation}
r_{\text{PA}} = \gamma \left( \frac{N_0}{N} - 1 \right) \text{ .}
\end{equation}
Using the 2\% value, this gives an estimated rate of $r_{\text{PA}} = 1.0 \times 10^{-2}$ s$^{-1}$ per atom, with lower and upper bounds of $6.8 \times 10^{-3}$ and $2.0 \times 10^{-2}$ s$^{-1}$ per atom, respectively. The total PA rate is $4.1 \times 10^5$ molecules per second, with bounds of $2.7 \times 10^5$ and $8.2 \times 10^5$  s$^{-1}$.

It should be noted that some trap loss is due to molecules decaying to free atoms, and if some of these are recaptured by the MOT we will underestimate the true PA rate. The fraction of $2 \, (0_g^+)$, $v' = 111$ molecules that decay to free atoms is 70\%, which could cause the PA rate to be higher by a factor of 3. If we use the high estimate for $P_{\text{ionization}}$, the resulting $3.1 \times 10^6$ molecules per second estimated from the ion counting rate is in only slight disagreement with the estimate based on trap loss and possible recapture.

This in turn is consistent with very strong coupling of the long-range $2 \, (1_g), \, v' = 155$ level and the short-range $2 \, (0_g^+), v' = 111, J' = 5$ level. Since molecules formed at short range appear to account for nearly all photoassociated molecules, the wavefunction must be strongly mixed.

For further comparison, we can calculate the molecule formation rate for the $2 \, (0_g^+), v' = 110, J' = 5$ level, which does not benefit from resonant coupling, using the same methods. In the spectrum of Fig. \ref{pa}(a), the $a \, ^3 \Sigma_u^+, v'' = 20$ level is detected. It has a FCF of $6.60 \times 10^{-2}$, and a peak ion signal size of 46 ions per shot. This gives a molecule production rate of $6.0 \times 10^5$ per second (with a range from $1.8 \times 10^5$ to $1.1 \times 10^6$), an order of magnitude less than the $1.1 \times 10^{7}$ molecules per second from the coupled level.

\section{Hyperfine Structure and Coupling}\label{structure}
\begin{figure}[tb]
\begin{center}
\includegraphics[width=\columnwidth]{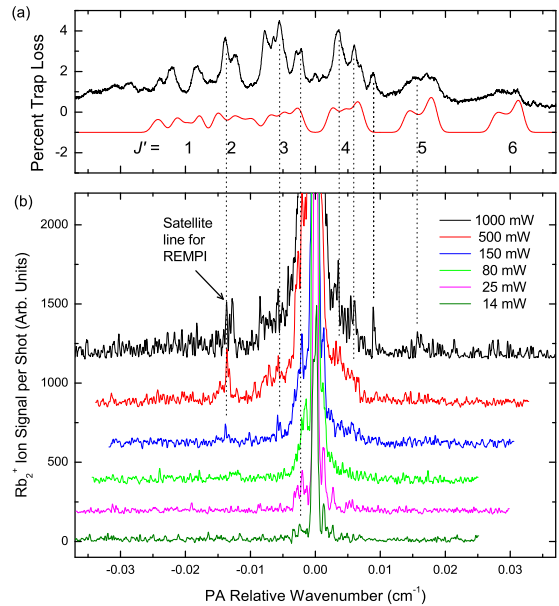}
\caption{(Color online) \textbf{(a)} Hyperfine structure of the $v' = 155$ level of the $2 \, (1_g)$ state, averaged from six trap loss scans. The energy zero point is the position of the $v' = 111$, $J' = 5$ level of the $2 \, (0_g^+)$ state. Below the data is a model fit using a hyperfine Hamiltonian from Ref.~\cite{bergeman12} and described in the text. The model is vertically offset from the experimental data, for clarity. \textbf{(b)} A series of scans at progressively lower PA intensities through the $v' = 111, \, J' = 5$ level of the $2 \, (0_g^+)$ state. The scans are vertically offset and smoothed for clarity. To the left of the main peak is the satellite feature whose REMPI spectrum is shown in Fig. \ref{rempi}. Both panel (a) and panel (b) are on the same scale, and referenced to the position of the $2 \, (0_g^+)$, $v' = 111$, $J' = 5$ level at $12798.17(3)$ cm$^{-1}$.}
\label{dollloss}
\end{center}
\end{figure}

To investigate the suspected hyperfine structure mentioned in Sec. \ref{experiment}, we undertook a series of high-resolution PA scans through the $J' = 5$ line. Each successive scan was done at a lower PA intensity, to better show features close to the central line. When aligned, as in Fig. \ref{dollloss}(b), they show significant, consistent structure around the central rotational level.  As indicated by dotted lines showing some of the strongest features, most of the satellite features correspond to trap loss of the $2 \, (1_g)$ state shown in Fig. \ref{dollloss}(a). This structure cannot be directly from decay products of the unperturbed $2 \, (1_g)$ state, as the FCFs for decay of $v' = 155$ to the $a \, ^3 \Sigma_u^+$ state are non-vanishing only for $v'' = 38$ and 39. These two levels have never been observed in REMPI in our apparatus, and are believed to be photodissociated quickly by the PA laser. In addition, the REMPI spectrum of the marked satellite feature (displayed in Fig. \ref{rempi}) clearly shows the level structure of deeply-bound $a \, ^3 \Sigma_u^+$ state molecules closely matching the spectrum of the strong central peak. Thus, as in the REMPI spectrum of the $J' = 5$ satellite feature in Fig. \ref{rempi}, this splitting appears to be rotational and hyperfine structure induced by coupling to the $v' = 155$ vibrational level of the $2 \, (1_g)$ state. It is also worth noting that the signal-to-noise ratio when using as little as 14 mW of PA power is still quite usable for spectroscopy, and is produced at a PA laser power far below the 500 mW to 1000 mW typically used in our experiment.

As discussed above, our trap loss scans of the $v' = 155$ level show significant rotational and hyperfine structure, though it is not well-resolved. We believe the linewidth is not a result of our scan speed, but is due to some combination of natural linewidth and broadening due to the high laser power needed to produce observable trap loss. An averaged scan, compiled from six high-resolution scans, is shown in Fig. \ref{dollloss}(a). This spectrum is difficult to assign \emph{a priori}, as the rotational and hyperfine level spacings overlap, particularly for low $J$ levels. In order to gain insight into this structure, we model these data with a simulated spectrum using a Hamiltonian described in Ref.~\cite{bergeman12}, 
\begin{align}
\mathcal{H} &= A_v \Omega i + \frac{\hbar^2 \vec{\ell}^2}{2 \mu R^2} \nonumber \\
&= A_v \left( i + \eta (i+3)^2 \right) + \frac{\hbar^2}{2 \mu R^2} \left( \vec{F} - \vec{J} - \vec{I} \right)^2 \text{ .}
\end{align}
This can be rewritten as
\begin{multline}
\mathcal{H} = A_v \left( i + \eta (i+3)^2 \right) + B_v \bigl( F(F+1) + 2\\ + I(I+1) - 2 \phi \Omega - F_+I_- - F_-I_+ - 2 \phi i + 2 \Omega i \bigr)
\end{multline}
where $A_v$ is the hyperfine coupling term, arising primarily from the $5 \, ^2 S$ Fermi contact interaction, $B_v$ is the rotational constant, $i$, $\phi$, and $\Omega$ are the projections of $\vec{I}$, $\vec{F}$, and $\vec{J}$ on the internuclear axis, respectively, and $\eta$ is a fitting constant for the term that is quadratic in $i$ (this term is added to allow for variation in $A_v$ with $i$). Using the parameters $A_v = 2.8 \times 10^{-6}$ cm$^{-1}$, $\eta = 8$, and $B_v = 0.00095$ cm$^{-1}$, the model qualitatively reproduces the observed spectrum, as is shown in Fig. \ref{dollloss}(a). One minor issue with this fit is that it requires a rotational constant $B_v$ that is smaller than the $B_v = 0.0012$ cm$^{-1}$ calculated from our potentials. The calculated $B_v$ values from other potentials have been relatively accurate, although they display a slight tendency to overestimate the $B_v$ values compared to experiment~\cite{bellos11, bellos12}. The model also does not include weighting of the incoming partial waves, which are affected by the thermal population of the MOT as well as the $l = 4$ shape resonance. Note that we have tried to model neither the hyperfine structure of the $v' = 111$ level of the $2 \, (0_g^+)$ state nor the coupling between the two hyperfine-split $J' = 5$ levels $\left( 2 \, (0_g^+) \sim 2 \, (1_g) \right)$.

From the spectrum in Fig. \ref{dollloss}(a), we measure the spacing between $v' = 111$, $J' = 5$ of the $2 \, (0_g^+)$ state and $v' = 155$, $J' = 5$ of the $2 \, (1_g)$ state to be $\Delta \approx 1.67 \times 10^{-2}$ cm$^{-1}$. We believe that levels other than $J' = 5$ of the $2 \, (0_g^+)$ state do not contribute, as the other $J$ levels are energetically much further away. If we know the unperturbed spacing of the levels and can measure a shift, we can estimate the strength of the coupling interaction via $\Delta = \sqrt{4 | H_{12} |^2 + \Delta_0^2}$, where $\Delta_0$ is the unperturbed spacing and $| H_{12} |$ is the interaction term of the Hamiltonian.

By fitting the rotational progression of $v' = 111$, $J' = 5$, using a simple $E_J = B_v J (J + 1)$ model both with and without the $J' = 5$ level, we find a shift of $7 \times 10^{-4}$ cm$^{-1}$. This is within the $1.3 \times 10^{-3}$ cm$^{-1}$ FWHM linewidth, but can still help establish an estimate of the coupling strength. Assuming the shift is symmetric with the other coupling partner, we find $| H_{12} | \approx 3.4 \times 10^{-3}$ cm$^{-1}$, an extremely small value.

Since the two states that are involved in this coupling have $\Omega = 0$ and $\Omega = 1$, the coupling must be the result of an inhomogeneous perturbation. There are two primary causes of such perturbations---the non-Born-Oppenheimer $S$-uncoupling and the $L$-uncoupling operators~\cite{field04}. The $S$-uncoupling operator has selection rules $\Delta S = 0$, $\Delta \Omega = \Delta \Sigma = \pm 1$, and $\Delta \Lambda = 0$. However, this operator only couples $\Omega$ components of the same electronic state multiplet. Additionally, our Hund's case (a) $2 \, ^1 \Sigma_g^+$ and $1 \, ^1 \Pi_g$ states have $\Lambda = 0$ and $\Lambda = 1$, respectively, ruling out an $S$-uncoupling mediated perturbation. The $L$-uncoupling operator has selection rules $\Delta \Omega = \Delta \Lambda = \pm 1$ and $\Delta S = 0$. This operator, having the form $\left( 1/2 \mu R^2 \right) \left(\mathbf{J^+ L^-} + \mathbf{J^- L^+} \right)$, can couple different electronic states. The observed resonant coupling is thus likely the result of an $L$-uncoupling mediated perturbation.

\section{Conclusion}
We have shown that there is resonant coupling between a pair of $J' = 5$ levels in the $2 \, ^1 \Sigma_g^+$ and $1 \, ^1 \Pi_g$ states. This coupling causes an order-of-magnitude increase in the production of $a \, ^3 \Sigma_u^+$ state molecules, compared with nearby vibrational levels of the $2 \, ^1 \Sigma_g^+$ state, and yields an approximate PA rate of $5.4 \times 10^{6}$ molecules per second. This coupling provides a strong pathway for creating deeply bound $a \, ^3 \Sigma_u^+$ state molecules. As it connects high-$v$ levels of the $a \, ^3 \Sigma_u^+$ state (through the potential of the long-range component) with more deeply bound levels (through the potential of the short-range component), it can also provide an experimental pathway for molecule transfer.

\begin{acknowledgments}
We gratefully acknowledge support from the NSF and AFOSR MURI. We also thank Tom Bergeman and Olivier Dulieu for helpful discussions.
\end{acknowledgments}

\bibliography{ultracold_references}

\end{document}